\documentclass[a4paper]{article}

\usepackage{INTERSPEECH2021}
\usepackage{cite}
\usepackage{url}
\usepackage{graphicx}
\usepackage{hyperref}
\usepackage{verbatim}
\usepackage{multirow}
\usepackage[table,xcdraw]{xcolor}
\usepackage{array}
\usepackage{tabularx}
\newcolumntype{Y}{>{\centering\arraybackslash}X}
\newcommand{\norm}[1]{\left\lVert#1\right\rVert}
\title{Relational Data Selection for Data Augmentation of Speaker-dependent Multi-band MelGAN Vocoder}
\name{Yi-Chiao Wu$^1$, Cheng-Hung Hu$^2$, Hung-Shin Lee$^2$, Yu-Huai Peng$^2$, Wen-Chin Huang$^1$, Yu Tsao$^2$, Hsin-Min Wang$^2$, and Tomoki Toda$^1$}

\address{
  $^1$Nagoya University, Japan\\
  $^2$Academia Sinica, Taiwan}
\email{yichiao.wu@g.sp.m.is.nagoya-u.ac.jp, tomoki@icts.nagoya-u.ac.jp}

\begin{document}

\maketitle
\begin{abstract}
Nowadays, neural vocoders can generate very high-fidelity speech when a bunch of training data is available. Although a speaker-dependent (SD) vocoder usually outperforms a speaker-independent (SI) vocoder, it is impractical to collect a large amount of data of a specific target speaker for most real-world applications. To tackle the problem of limited target data, a data augmentation method based on speaker representation and similarity measurement of speaker verification is proposed in this paper. The proposed method selects utterances that have similar speaker identity to the target speaker from an external corpus, and then combines the selected utterances with the limited target data for SD vocoder adaptation. The evaluation results show that, compared with the vocoder adapted using only limited target data, the vocoder adapted using augmented data improves both the quality and similarity of synthesized speech.   
\end{abstract}

\section{Introduction}
A vocoder~\cite{vocoder_1939} is a speech codec to analyze speech into acoustic features and synthesize the acoustic features back to speech. Conventional vocoders~\cite{straight, world} built following the source-filter model~\cite{source_filter} have been widely used in speech synthesis tasks. However, the quality of the synthetic speech is usually degraded because the phase information and temporal details are discarded during the analysis-synthesis process of the conventional vocoders. Many neural speech generation models~\cite{wavenet, samplernn, wavernn, pwn, clarinet, waveglow, flowavenet, pwg, melgan, gantts} have been proposed to directly model speech waveforms without many ad~hoc assumptions of speech generation. Using these models as vocoders~\cite{sd_wn_vocoder, si_wn_vocoder} to synthesize speech based on the acoustic features extracted by conventional vocoders also remarkably improve the naturalness of the synthetic speech.

Although a speaker-dependent (SD) model usually outperforms a speaker-independent (SI) model~\cite{ sd_wn_vocoder, si_wn_vocoder }, collecting much data from a user is impractical. An efficient way to develop an SD vocoder is to first train a multi-speaker vocoder using a multi-speaker corpus, and then adapt the multi-speaker vocoder to the SD vocoder using the few-shot target data. However, this method still requires about five minutes of target data to develop a stable SD vocoder~\cite{ nu_np_2018, nu_p_2018}. On the other hand, an SI vocoder trained with a varied corpus may outperform an SD vocoder trained with a relatively small corpus in terms of speech quality~\cite{si-vocoder}. However, the difference of speaker similarity has not been well investigated. In this paper, we first explore a more challenging scenario, i.e., one-shot adaptation, where the available target data is around 30s. Then, the speaker similarity difference between SI and SD vocoders is investigated.  

The performance of most neural models is highly correlated with the amount and diversity of training data due to the data-driven nature. Therefore, the use of generative models to generate augmented data is straightforward and has been applied to many recognition tasks. For example, well-trained text-to-speech (TTS) systems have been used to generate augmented data for training automatic speech recognition (ASR)~\cite{ttsasr_2018, ttsasr_2019, ttsasr_2020} and speech translation~\cite{ttsmt_2019} systems. A vocoder has been used to generate augmented data with a variety of fundamental frequency ($F_0$) patterns to train an $F_0$ estimator~\cite{vocoderf0_2019}, and a voice conversion (VC) framework has been used to generate unseen speaker data to train a speaker embedding extractor~\cite{vcsv_2019}. Even generation tasks, such as TTS~\cite{ttsbytts}, can benefit from using augmented data generated by another TTS system. However, there is still a performance gap between the models trained with sufficient natural data and augmented synthetic data~\cite{ttsasr_2019}.

In this paper, different from using generative models to produce augmented data, since leveraging natural data may avoid error propagation and keep the entire framework simple, we propose a data selection method to select augmented utterances from an external corpus. The speaker representation and similarity measurement from speaker verification (SV)~\cite{xvector} are used to formulate the selection criteria, and the selected utterances are presumed to have similar speaker identities to the target speaker. An SI vocoder is first trained with the multi-speaker external corpus, and then the one-shot target data and selected augmented utterances are used together to adapt the SI vocoder to the SD vocoder. Multi-band MelGAN~\cite{multi-melgan} is adopted as the neural vocoder because of its light architecture. According to our evaluation results, the vocoder adapted with one-shot target data and augmented data achieves higher quality and similarity of synthesized speech compared with the vocoder adapted with only one-shot target data. To our best knowledge, our method is the first approach to train SD neural vocoders by using SV technology for data augmentation from an external corpus.

\section{Baseline multi-band MelGAN vocoder}
In this section, we introduce the baseline multi-band MelGAN vocoder, which is a convolutional neural network (CNN)-based non-autoregressive (non-AR) raw waveform generation model.

\subsection{MelGAN with a multi-resolution STFT loss}
A classic generative adversarial net (GAN)~\cite{gan} architecture is adopted in MelGAN to convert the input mel-spectrogram into speech samples. Instead of one discriminator, MelGAN utilizes $K$ discriminators ($D_k$) running at different temporal resolutions to capture the hierarchical structure of speech. Given a generator ($G$), natural speech $\boldsymbol{y}$, and the input mel-spectrogram $\boldsymbol{s}$, the discriminator adversarial loss ($L_{\mathrm{D}}$) is formulated as
\begin{align}
&L_{\mathrm{D}}(G, D) \nonumber \\
&=\displaystyle\sum\limits_{k=1}^K \frac{1}{K} (\mathbb{E}_{\boldsymbol{y}}\left[(1-D_{k}(\boldsymbol{y}))^{2}\right]
+\mathbb{E}_{\boldsymbol{s}}\left[D_{k}(G(\boldsymbol{s}))^{2}\right]),
\label{eq:ld}
\end{align}
where $k$ is the discriminator index.

Moreover, the generator loss ($L_{\mathrm{G}}$) includes an auxiliary loss in addition to the original adversarial loss ($L_{\mathrm{adv}}$). Specifically, to improve the stability of training, vanilla MelGAN adopts a feature matching loss to regularize the discriminator intermediate feature maps of real and fake data. However, since the multi-resolution short-time Fourier transform (STFT) loss~\cite{pwg} is more meaningful and can make the training process converge fast~\cite{multi-melgan}, the feature matching loss is replaced by the multi-resolution STFT loss ($L_{\mathrm{sp}}$) when training our MelGAN vocoders. The generator loss is formulated as
\begin{align}
L_{\mathrm{G}}(G, D)=\lambda_{\mathrm{adv}} L_{\mathrm{adv}}(G, D)+L_{\mathrm{sp}}(G),
\label{eq:lg}
\end{align}
where $\lambda_{\mathrm{adv}}$ is a balance weight, which is set to 2.5 in this work. The $ L_{\mathrm{adv}}$ loss is formulated as
\begin{align}
L_{\mathrm{adv}}(G, D)=\mathbb{E}_{\boldsymbol{s}}\left[(1-D(G(\boldsymbol{s})))^{2}\right].
\label{eq:ladv}
\end{align} 
The $L_{\mathrm{sp}}$ loss is calculated based on the STFT features extracted using three setting groups, including the FFT size of [1024, 2048, 4096], the frame shift of [120, 240, 480], and the window length of [600, 1200, 2400]. 

The generator and discriminators of MelGAN are fully convolutional networks. The generator adopts several transposed CNN layers with residual networks and dilated CNN layers~\cite{dcnn} to gradually upsample the input mel-spectrogram to match the temporal resolution of the output waveform samples. A LeakyReLU~\cite{leakyrelu} activation is adopted following each CNN layer except the last output layer, which uses a tanh function to output waveforms. The multi-scale discriminators have an identical network structure but different downsampling factors. Downsampling is implemented using stride average pooling. More details can be found in the open source repository\footnotemark.

\subsection{Multi-band approach}
Directly modeling speech samples with a high sampling frequency ($f_s$) is challenging because of the speech signal has a high temporal resolution with a very long-term dependency, which usually result in the consumption of time and computing resources in the generation process. Decomposing the speech signal into several subbands can significantly improve the generation efficiency, because each subband signal is generated in parallel using a single network. The multi-band approach has been successfully applied to many AR~\cite{multi-wn-vocoder, multi-fftnet, multi-wavernn} and non-AR~\cite{multi-melgan} neural vocoders.

The $f_s$ of the speech signal processed in this paper is 44.1~KHz. The analysis and synthesis filters in~\cite{multi-wavernn, multi-melgan} are used to decompose the speech signal into five frequency bands, and the fullband signal is restored on the basis of the subband signals. The generator is trained to generate the subband signals in parallel. The inputs of the discriminators are the restored fullband signal. To improve the stability of multi-band training, the multi-resolution STFT loss is adopted for both fullband and subband signals. The setting groups of subband STFT include the FFT size of [384, 683, 171], the frame shift of [30, 60, 10], and the window length of [150, 300, 60]. More details can be found in~\cite{multi-wavernn, multi-melgan} and the open source repository\footnotemark[\value{footnote}].

\footnotetext{https://github.com/kan-bayashi/ParallelWaveGAN}

\section{Relational data selection}
To effectively develop the SD vocoder when very limited (one-shot) target data is available, we propose a data augmentation framework leveraging an external corpus (candidate pool) for adapting the SI vocoder to the target SD vocoder. A hierarchical framework based on data relationships is used to select suitable data for speaker adaptation. Three levels of relationships are considered. First, the speaker similarity is measured by the inter-speaker relationship between the target speaker and the candidate speaker. Second, the inter-candidate-speaker relationships are established to verify the speaker-wise confidence. Third, the reliability of each candidate utterance is regularized by the relationship within the speaker.

\subsection{Speaker similarity}
Selecting external utterances with similar speaker identities to the target speaker for speaker adaptation is straightforward. Identity modeling and similarity measurement in SV technology can be used in this work. For identity modeling, we use the state-of-the-art x-vector~\cite{xvector} speaker representation. For similarity measurement, we use the probabilistic linear discriminant analysis (PLDA) ~\cite{plda_2006, plda_2007}. X-vector is a speaker embedding extracted from the intermediate layer of a speaker-identification neural network, and PLDA is a distance measurement designed to be generalized for arbitrary unseen classes. Therefore, the first selection criterion is formulated as
\begin{align}
\mathrm{PLDA}(\boldsymbol{x}_{n, i},\boldsymbol{x}_{Target}),
\label{eq:dc1}
\end{align}
where $\boldsymbol{x}$ denotes x-vector, $n$ is the candidate speaker index, $i$ is the utterance index, and $\boldsymbol{x}_{Target}$ is the average x-vector of all target utterances. The higher the PLDA score, the higher the speaker similarity.

\subsection{Speaker-wise divergence regularization}
Since a robust speaker embedding should be almost independent of the phonetic content, we assume that the utterance-wise x-vectors from the same speaker should be similar. According to the assumption, if the distribution of the x-vectors of a speaker is diverse, the reliability of these x-vectors may be low. To model the speaker-wise confidence within the candidate pool, an SD term is introduced to regularize the PLDA score. First, temperature sigmoid is applied to make all PLDA scores have the same sign,
\begin{align}
\mathrm{PLDA}'(\cdot) = \frac{1}{1+0.5\times e^{-\mathrm{PLDA}(\cdot)}}.
\label{eq:tsig}
\end{align}
Then, the second selection criterion is formulated as
\begin{align}
\frac{\mathrm{PLDA}'(\boldsymbol{x}_{n, i},\boldsymbol{x}_{Target})}{(\sigma_n)^\alpha},
\label{eq:dc2}
\end{align}
where $\sigma_n$ is the square root of the average squared Euclidean distance between each utterance x-vector and the mean x-vector of speaker $n$, and the weight $\alpha$ is set to 0.1 in this paper. High x-vector diversity of a speaker results in low speaker confidence.

\subsection{Utterance-wise divergence regularization}
Following the above assumption, the internal speaker relationship is introduced to tackle the outlier utterances within each speaker. That is, if the x-vector of an utterance is very different from the x-vectors of other utterances of the same speaker, the utterance is considered an outlier, and its x-vector is unreliable. Therefore, the denominator in Eq. (\ref{eq:dc2}) can be combined with an utterance-wise regularizer to evaluate the utterance reliability, and the third selection criterion is formulated as 
\begin{align}
\frac{\mathrm{PLDA}'(\boldsymbol{x}_{n, i},\boldsymbol{x}_{Target})}{(\sigma_n\lVert \boldsymbol{x}_{n, i}-\boldsymbol{u}_{n} \rVert_2)^\alpha},
\label{eq:dc3}
\end{align}
where $\norm{\cdot}_{2}$ denotes the Euclidean distance (L2 norm), and $\boldsymbol{u}_{n}$ is the mean x-vector of speaker $n$. The larger the Euclidean distance, the lower the reliability. 

In summary, the criteria in this section model different relationships among the target speaker and the individual speakers and utterances in the candidate pool. Each subsequent criterion is derived from the previous criterion.

\section{Experiments}

\subsection{Corpus}
The AIShell-3 and TST Mandarin corpora provided by the ICASSP2021-M2VoC organizer~\cite{m2voc2021} were used in the experiments. The training set of AIShell-3, which includes 137 female and 37 male speakers, was used to train the baseline SI vocoder and was used as the candidate pool. The female speakers have 50,117 utterances ($\sim$50~hours) in total, and the male speakers have 13,145 utterances ($\sim$13~hours) in total. One female and two male speakers in the Track~1 subset of TST were used as the target speakers. Each target speaker has 90 training utterances (6--10~mins) and 10 testing utterances. To simulate the one-shot scenario, the first five training utterances ($\sim$30~s) of each target speaker were used as the limited target data. The $f_s$ and bit-depth of all utterances were set to 44.1~KHz and 16.

\subsection{Experimental setting}
The mel-spectrogram of 80 mel-filter banks was used as the input of the vocoder. The hop size was 220 samples, and the FFT size was 2048. The pre-trained models of the SITW (speakers in the wild) x-vector system\footnotemark was used for extraction of 512-dimensional x-vectors and calculation of PLDA scores. All input audio files were downsampled to 16~KHz to match the working $f_s$ of these speaker models. The x-vector of each candidate speaker was the average x-vector of all the utterances from that speaker, and the x-vector of each target speaker was also the average x-vector of the available utterances corresponding to different scenarios.

\subsection{Model description}
Six multi-band MelGAN vocoders were evaluated. That is, an SI (multi-speaker) vocoder was first trained using the training set of AIShell-3, and then the adapt5 and adapt90 SD vocoders were developed by adapting the SI vocoder using five and 90 target utterances, respectively. The adapt90 vocoder was taken as an upper bound in this section. For the proposed vocoders, to match the amount of adaptation data of adapt90, 85 utterances were selected from the AIShell-3 training set using the proposed criteria 1--3, and then combined with the five target utterances to form the adaptation sets DC1--DC3, where DC denotes the data selection criterion. The proposed SD vocoders were adapted from the SI vocoder using the DC1--DC3 sets, respectively.

The SI vocoder was trained for 1M iterations, and the discriminators were jointly trained with the generator from the 200,000-th iteration. Adam optimizer~\cite{adam} was used, and the learning rate was set to 0.001 without decay. During speaker adaptation, the discriminators and generator were updated. The adaptation iteration number for the adapt5 vocoders was set to 1000, and the iteration number for the adapt90 and DC1--DC3 vocoders was set to 9000. It was difficult to find the optimal number of iterations for the adapt5 vocoders, because the adaptation data was very limited. Therefore, the iteration number of 1000 was a compromise choice for the three target speakers.

\subsection{Objective evaluation}
\begin{table}[t]
\caption{Objective evaluation results.}
\vspace{-3.5mm}
\label{tb:objective}
\fontsize{8pt}{9.6pt}
\selectfont
{%
\begin{tabularx}{\columnwidth}{@{}p{1.3cm}YYYYYY@{}}%Y means centering, X for left
\toprule
         & SI      & Adapt90           & Adapt5  & DC1   & DC2     & DC3     \\ \midrule
LSD (dB) & 1.07    & \textbf{1.00}     & 1.08    & 1.09  & 1.08    & 1.04    \\
MCD (dB) & 6.36    & \textbf{5.43}     & 5.96    & 6.06  & 6.03    & 6.08    \\ 
$F_0$ (Hz)  & 70.8 & \textbf{67.3}     & 69.5    & 68.2  & 67.9    & 70.0    \\
$U/V$ (\%)  & 15.6 & \textbf{13.7}     & 14.9    & 15.1  & 14.9    & 14.7    \\
PLDA     & 33.2    & \textbf{39.5}     & 29.0    & 34.8  & 32.6    & 33.6    \\
CosSim   & 0.82    & \textbf{0.88}     & 0.81    & 0.84  & 0.84    & 0.84    \\ \bottomrule
\end{tabularx}%
}
\vspace{-6mm}
\end{table}
Objective evaluations based on spectral accuracy, source excitation accuracy, and speaker similarity were conducted. The spectral accuracy was evaluated in terms of log spectral distortion (LSD) and mel-cepstral distortion (MCD). For source excitation accuracy, we measured the root mean square error (RMSE) of $F_0$ and $U/V$ decision error. For speaker similarity measurement, the PLDA score and cosine similarity (CosSim) of two x-vectors were used.
Mel-cepstrum ($mcep$), $F_0$, and unvoiced/voicded ($U/V$) features were extracted using WORLD~\cite{world}. The ground truth acoustic features and x-vectors were extracted from the natural testing utterances. 

The average evaluation results of the three target speakers are shown in Table~\ref{tb:objective}. As expected, the adapt90 vocoders achieve the best performance in all metrics, which shows the effectiveness of speaker adaptation when the adaptation data is relatively sufficient. In contrast, the performance of the adapt5 vocoders is much worse than that of the adapt90 vocoders. This is due to the instability and quality degradation caused by much less adaptation data.

\footnotetext{ https://kaldi-asr.org/models/m8}

For spectral accuracy evaluation, we can see that the adapt5 vocoders were better than the SI vocoder in MCD, but worse in LSD. One possible reason is that $mcep$ is dominated by spectral envelope components in low-frequency bands, and due to improved SD component modeling, the adapt5 vocoders can achieve higher formant modeling accuracy. When listening to the utterances produced by adapt5, we could perceive a similar trend. That is, despite the slight improvement in the similarity of timbre, the speech generated by adapt5 suffered from severe musical noise and oversmoothing. The musical noise and oversmoothing effects may not be well modeled by $mcep$, but will be reflected in the LSD measurement. Moreover, the spectral accuracy results show that the the proposed methods are effective because the DC* vocoders achieve lower MCD and similar/lower LSD than the SI vocoder. Compared with the adapt5 vocoders, the DC3 vocoders achieve lower LSD and slightly higher MCD, which implies that external data may slightly reduce the accuracy of SD component modeling. 

Since both the voiced and unvoiced parts were involved in the RMSE calculation of $F_0$, the error of $F_0$ is correlated with the $U/V$ error. Generally speaking, prosodic characteristics are highly related to speaker identity. As shown in Table~\ref{tb:objective}, the SI vocoder yielded higher $F_0$ and $U/V$ errors than the SD vocoders. The results confirm the assumption and indicate that prosodic modeling can be improved by speaker adaptation. Moreover, the DC* vocoders achieved similar or lower prosodic errors to the adapt5 vocoders, indicating the potential of the proposed methods to further improve prosodic modeling.

For speaker similarity evaluation, the DC* vocoders achieved higher PLDA and CosSim scores than the SI and adapt5 vocoders, which shows the effectiveness of the proposed method in improving the speaker similarity of synthetic speech. Again, one possible reason for the worst performance of the adapt5 vocoders is that the musical noise and oversmoothing can reduce the speaker similarity. In addition, since DC2 and DC3 apply regularizations to the PLDA score, it is reasonable that the DC1 vocoders are slightly better than the DC2 and DC3 vocoders in the PLDA score, but CosSim remains the same. 
%More discussion will be presented in Section~\ref{discussion}.

\subsection{Subjective evaluation}
\begin{table}[t]
\caption{Subjective evaluation results (MOS values).}
\vspace{-3.5mm}
\label{tb:subjective}
\fontsize{8pt}{9.6pt}
\selectfont
{%
\begin{tabularx}{\columnwidth}{@{}p{0.8cm}YYYYY@{}}
\toprule
           & Natural      & SI           & Adapt90      & Adapt5       & DC3   \\ \midrule
Quality    & 4.96$\pm$.03 & 3.47$\pm$.14 & \textbf{3.71}$\pm$.13 & 2.65$\pm$.14 & 3.29$\pm$.13  \\
Similarity & -            & 3.57$\pm$.16 & \textbf{4.23}$\pm$.14 & 3.43$\pm$.16 & 3.71$\pm$.15 \\ \bottomrule
\end{tabularx}%
}
\vspace{1mm}
\caption{Statistics of the selected utterances (female target).}
\vspace{-3.5mm}
\label{tb:statistic}
\fontsize{8pt}{9.6pt}
\selectfont
{%
\begin{tabularx}{\columnwidth}{@{}p{0.4cm}YYYY@{}}
\toprule
 & Number of speakers & Suspected utterances & Utterance overlap (\%)& Speaker overlap (\%)\\ \midrule
DC1 & 18 & 9 & reference & reference \\
DC2 & 7  & 2 & 15.3 & 32.0 \\
DC3 & 16 & 5 & 10.6 & 70.6 \\ \bottomrule
\end{tabularx}%
}
\vspace{-6mm}
\end{table}
Two mean opinion score (MOS) evaluations were conducted for speech quality and speaker similarity, respectively. In the quality evaluation, each utterance was given a score in the range 1--5 by a listener to evaluate the naturalness. In the similarity evaluation, the listener listened to a pair of natural and synthetic utterances at a time, and gave a score between 1 and 5 to evaluate the speaker similarity of the synthetic utterance to the natural utterance. In both evaluations, the higher the score, the better the performance. Nine native speakers participated in the evaluation using the same device. Five types of speech were compared, including natural speech and the speech produced by the SI, adapt5, adapt90, and DC3 vocoders. For each of the three target speakers, there were 10 natural utterances and 40 synthetic utterances produced by four vocoders. The 150 utterances were divided into two subsets, and each subset was evaluated by at least five listeners. Demo samples can be found on
our website~\cite{demo}.

As shown in Table~\ref{tb:subjective}, the superior performance of the adapt90 vocoder in both quality and similarity measurements proves the importance of speaker adaptation to the vocoder. However, due to severe musical noise and oversmoothing, the adapt5 vocoder achieved the worst naturalness and similarity. It even gave worse performance than the SI vocoder. The results show that one-shot speaker adaptation is challenging, and the vocoders adapted with extremely limited data tend to be unstable and cannot be generalized. However, the proposed DC3 vocoder significantly outperformed the adapt5 vocoder in both quality and similarity measurements. The results confirm the effectiveness of the proposed data augmentation method derived from SV technology for the speaker adaptation of SD vocoders.

Although the DC3 vocoder is superior to the SI vocoder in terms of speaker similarity, the SI vocoder is robust to unseen speakers in terms of quality. The result is reasonable because the SI vocoder is trained with a large amount of data from many speakers. The great diversity of the training data allows the SI vocoder to generate stable speech for unseen speakers, even if the similarity is still insufficient. Therefore, we may conclude that the quality of synthetic speech is highly related to the amount of training/adaptation data, and the proposed data augmentation method can improve the speaker similarity of the SI vocoder even when only 30s target data is available. Moreover, according to the results, there is a significant quality gap between natural and synthetic speech, which implies that the current multi-band MelGAN vocoder may not be able to handle high $f_s$ speech generation. There is still room for improving neural vocoders to generate signals with a high $f_s$.

\subsection{Discussion} \label{discussion}
\begin{figure}[t]
\begin{center}
\includegraphics[width=0.99\columnwidth]{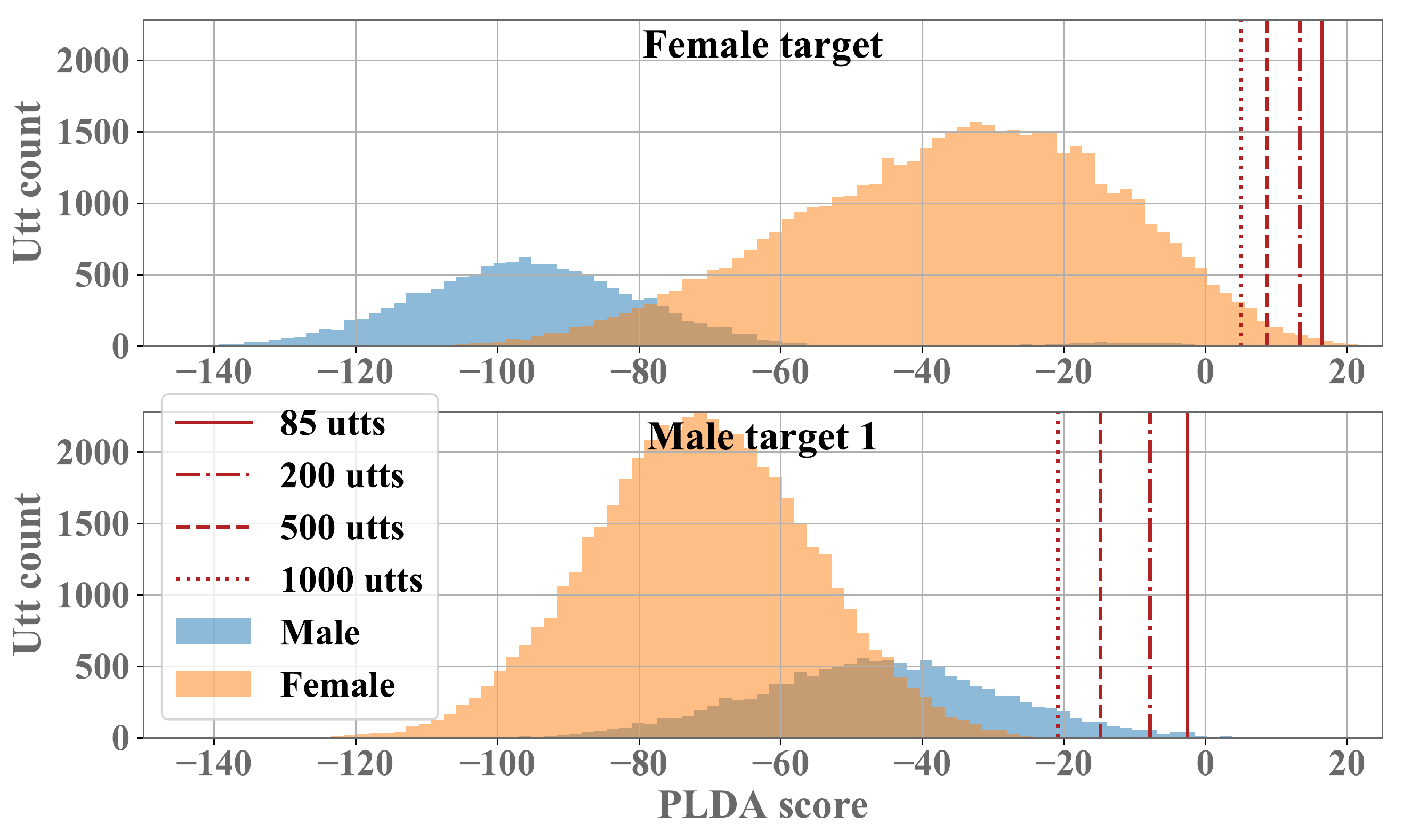}
\vspace{-5mm}
\caption{PLDA score distributions of candidate utterances.}
\label{fig:plda}
\end{center}
\vspace{-10mm}
\end{figure}
The score distributions for the female and the first male targets in Fig.~\ref{fig:plda} show the PLDA scores reflect the high correlation between gender and speaker similarity as expected. The selection thresholds of different utterance numbers imply the challenge to optimize the selection numbers because of the rapid PLDA score decrease for the increased selected utterances.

Since the female pool is much larger, we report the statistics of the selection utterances for the female target using DC1--3 in Table~\ref{tb:statistic}, and the results of DC1 are taken as the reference. The speaker regularizer is adopted to filter the utterances of low confidence speakers, and the speaker number of the selected utterances does decrease in DC2. An utterance might be an outlier if it is the only utterance from a speaker to be selected, and we call it a suspected utterance. The utterance regularizer is utilized to remove the outlier utterances, and the suspected utterance number does reduce in DC3. 

The low speaker overlap rate between the DC1 and DC2 sets means the speaker regularizer filters many speakers. The higher speaker but lower utterance overlap rates between the DC1 and DC3 sets show that the utterance regularizer makes DC3 select more representative utterances of each speaker, and the influence of the speaker regularization is eased in DC3. 

\section{Conclusions}
In this paper, we first investigated the influence of speaker adaptation and adaptive data volume on the effectiveness of SD neural vocoder. Then, we proposed selection criteria based on SV technology to leverage an external corpus for improving speaker adaptation. The proposed criteria explored the different relationships in the data space. The evaluation results show the effectiveness of the proposed framework in one-shot speech synthesis. For future work, the amount of selected data and the weight of regularizers should be further optimized.

\section{Acknowledgements}

This work was supported by JSPS KAKENHI Grant Number 17H06101 and JST, CREST Grant Number JPMJCR19A3.

\bibliographystyle{IEEEtran}

\bibliography{mybib}

% Generated by IEEEtran.bst, version: 1.13 (2008/09/30)
\begin{thebibliography}{10}
\providecommand{\url}[1]{#1}
\csname url@samestyle\endcsname
\providecommand{\newblock}{\relax}
\providecommand{\bibinfo}[2]{#2}
\providecommand{\BIBentrySTDinterwordspacing}{\spaceskip=0pt\relax}
\providecommand{\BIBentryALTinterwordstretchfactor}{4}
\providecommand{\BIBentryALTinterwordspacing}{\spaceskip=\fontdimen2\font plus
\BIBentryALTinterwordstretchfactor\fontdimen3\font minus
  \fontdimen4\font\relax}
\providecommand{\BIBforeignlanguage}[2]{{%
\expandafter\ifx\csname l@#1\endcsname\relax
\typeout{** WARNING: IEEEtran.bst: No hyphenation pattern has been}%
\typeout{** loaded for the language `#1'. Using the pattern for}%
\typeout{** the default language instead.}%
\else
\language=\csname l@#1\endcsname
\fi
#2}}
\providecommand{\BIBdecl}{\relax}
\BIBdecl

\bibitem{vocoder_1939}
H.~Dudley, ``Remaking speech,'' \emph{The Journal of the Acoustical Society of
  America}, vol.~11, no.~2, pp. 169--177, 1939.

\bibitem{straight}
H.~Kawahara, I.~Masuda-Katsuse, and A.~De~Cheveigne, ``Restructuring speech
  representations using a pitch-adaptive time--frequency smoothing and an
  instantaneous-frequency-based {F0} extraction: Possible role of a repetitive
  structure in sounds,'' \emph{Speech Communication}, vol.~27, no. 3-4, pp.
  187--207, 1999.

\bibitem{world}
M.~Morise, F.~Yokomori, and K.~Ozawa, ``{WORLD}: a vocoder-based high-quality
  speech synthesis system for real-time applications,'' \emph{IEICE
  Transactions on Information and Systems}, vol.~99, no.~7, pp. 1877--1884,
  2016.

\bibitem{source_filter}
R.~McAulay and T.~Quatieri, ``Speech analysis/synthesis based on a sinusoidal
  representation,'' \emph{IEEE Transactions on Acoustics, Speech, and Signal
  Processing}, vol.~34, no.~4, pp. 744--754, 1986.

\bibitem{wavenet}
A.~van~den Oord, S.~Dieleman, H.~Zen, K.~Simonyan, O.~Vinyals, A.~Graves,
  N.~Kalchbrenner, A.~Senior, and K.~Kavukcuoglu, ``Wave{N}et: A generative
  model for raw audio,'' in \emph{Proc. SSW9}, Sept. 2016, p. 125.

\bibitem{samplernn}
S.~Mehri, K.~Kumar, I.~Gulrajani, R.~Kumar, S.~Jain, J.~Sotelo, A.~Courville,
  and Y.~Bengio, ``Sample{RNN}: {A}n unconditional end-to-end neural audio
  generation model,'' in \emph{Proc. ICLR}, Apr. 2017.

\bibitem{wavernn}
N.~Kalchbrenner, E.~Elsen, K.~Simonyan, S.~Noury, N.~Casagrande, E.~Lockhart,
  F.~Stimberg, A.~van~den Oord, S.~Dieleman, and K.~Kavukcuoglu, ``Efficient
  neural audio synthesis,'' in \emph{Proc. ICML}, July 2018, pp. 2415--2424.

\bibitem{pwn}
A.~van~den Oord, Y.~Li, I.~Babuschkin, K.~Simonyan, O.~Vinyals, K.~Kavukcuoglu,
  G.~van~den Driessche, E.~Lockhart, L.~C. Cobo, F.~Stimberg, N.~Casagrande,
  D.~Grewe, S.~Noury, S.~Dieleman, E.~Elsen, N.~Kalchbrenner, H.~Zen,
  A.~Graves, H.~King, T.~Walters, D.~Belov, and D.~Hassabis, ``Parallel
  {W}ave{N}et: Fast high-fidelity speech synthesis,'' in \emph{Proc. ICML},
  July 2018, pp. 3915--3923.

\bibitem{clarinet}
W.~Ping, K.~Peng, and J.~Chen, ``Clari{N}et: Parallel wave generation in
  end-to-end text-to-speech,'' in \emph{Proc. ICLR}, May 2019.

\bibitem{waveglow}
R.~Prenger, R.~Valle, and B.~Catanzaro, ``Wave{G}low: A flow-based generative
  network for speech synthesis,'' in \emph{Proc. ICASSP}, May 2019, pp.
  3617--3621.

\bibitem{flowavenet}
S.~Kim, S.-G. Lee, J.~Song, J.~Kim, and S.~Yoon, ``{F}lo{W}ave{N}et : A
  generative flow for raw audio,'' in \emph{Proc. ICML}, June 2019, pp.
  3370--3378.

\bibitem{pwg}
R.~Yamamoto, E.~Song, and J.-M. Kim, ``Parallel {W}ave{GAN}: A fast waveform
  generation model based on generative adversarial networks with
  multi-resolution spectrogram,'' in \emph{Proc. ICASSP}, May 2020, pp.
  6199--6203.

\bibitem{melgan}
K.~Kumar, R.~Kumar, T.~de~Boissiere, L.~Gestin, W.~Z. Teoh, J.~Sotelo,
  A.~de~Br\'{e}bisson, Y.~Bengio, and A.~C. Courville, ``Mel{GAN}: Generative
  adversarial networks for conditional waveform synthesis,'' in \emph{Proc.
  NeurIPS}, Dec. 2019, pp. 14\,910--14\,921.

\bibitem{gantts}
M.~Bi{\'n}kowski, J.~Donahue, S.~Dieleman, A.~Clark, E.~Elsen, N.~Casagrande,
  L.~C. Cobo, and K.~Simonyan, ``High fidelity speech synthesis with
  adversarial networks,'' in \emph{Proc. ICLR}, Apr. 2020.

\bibitem{sd_wn_vocoder}
A.~Tamamori, T.~Hayashi, K.~Kobayashi, K.~Takeda, and T.~Toda,
  ``Speaker-dependent {W}ave{N}et vocoder,'' in \emph{Proc. INTERSPEECH}, Aug.
  2017, pp. 1118--1122.

\bibitem{si_wn_vocoder}
T.~Hayashi, A.~Tamamori, K.~Kobayashi, K.~Takeda, and T.~Toda, ``An
  investigation of multi-speaker training for {W}ave{N}et vocoder,'' in
  \emph{Proc. ASRU}, Dec. 2017, pp. 712--718.

\bibitem{nu_np_2018}
Y.-C. Wu, P.~L. Tobing, T.~Hayashi, K.~Kobayashi, and T.~Toda, ``The {NU}
  non-parallel voice conversion system for the {V}oice {C}onversion {C}hallenge
  2018,'' in \emph{Proc. Odyssey}, June 2018, pp. 211--218.

\bibitem{nu_p_2018}
P.~L. Tobing, Y.-C. Wu, T.~Hayashi, K.~Kobayashi, and T.~Toda, ``{NU} voice
  conversion system for the {V}oice {C}onversion {C}hallenge 2018,'' in
  \emph{Proc. Odyssey}, June 2018, pp. 219--226.

\bibitem{si-vocoder}
\BIBentryALTinterwordspacing
J.~Lorenzo-Trueba, T.~Drugman, J.~Latorre, T.~Merritt, B.~Putrycz,
  R.~Barra-Chicote, A.~Moinet, and V.~Aggarwal, ``{T}owards achieving robust
  universal neural vocoding,'' in \emph{Proc. Interspeech 2019}, 2019, pp.
  181--185. [Online]. Available:
  \url{http://dx.doi.org/10.21437/Interspeech.2019-1424}
\BIBentrySTDinterwordspacing

\bibitem{ttsasr_2018}
\BIBentryALTinterwordspacing
A.~Tjandra, S.~Sakti, and S.~Nakamura, ``Machine speech chain with one-shot
  speaker adaptation,'' in \emph{Proc. Interspeech 2018}, 2018, pp. 887--891.
  [Online]. Available: \url{http://dx.doi.org/10.21437/Interspeech.2018-1558}
\BIBentrySTDinterwordspacing

\bibitem{ttsasr_2019}
A.~Rosenberg, Y.~Zhang, B.~Ramabhadran, Y.~Jia, P.~Moreno, Y.~Wu, and Z.~Wu,
  ``Speech recognition with augmented synthesized speech,'' in
  \emph{ASRU}.\hskip 1em plus 0.5em minus 0.4em\relax IEEE, 2019, pp.
  996--1002.

\bibitem{ttsasr_2020}
A.~Laptev, R.~Korostik, A.~Svischev, A.~Andrusenko, I.~Medennikov, and
  S.~Rybin, ``You do not need more data: improving end-to-end speech
  recognition by text-to-speech data augmentation,'' in \emph{CISP-BMEI}.\hskip
  1em plus 0.5em minus 0.4em\relax IEEE, 2020, pp. 439--444.

\bibitem{ttsmt_2019}
Y.~Jia, M.~Johnson, W.~Macherey, R.~J. Weiss, Y.~Cao, C.-C. Chiu, N.~Ari,
  S.~Laurenzo, and Y.~Wu, ``Leveraging weakly supervised data to improve
  end-to-end speech-to-text translation,'' in \emph{ICASSP}.\hskip 1em plus
  0.5em minus 0.4em\relax IEEE, 2019, pp. 7180--7184.

\bibitem{vocoderf0_2019}
M.~Airaksinen, L.~Juvela, P.~Alku, and O.~R{\"a}s{\"a}nen, ``Data augmentation
  strategies for neural network f0 estimation,'' in \emph{ICASSP}.\hskip 1em
  plus 0.5em minus 0.4em\relax IEEE, 2019, pp. 6485--6489.

\bibitem{vcsv_2019}
\BIBentryALTinterwordspacing
H.~Yamamoto, K.~A. Lee, K.~Okabe, and T.~Koshinaka, ``{S}peaker augmentation
  and bandwidth extension for deep speaker embedding,'' in \emph{Proc.
  Interspeech 2019}, 2019, pp. 406--410. [Online]. Available:
  \url{http://dx.doi.org/10.21437/Interspeech.2019-1508}
\BIBentrySTDinterwordspacing

\bibitem{ttsbytts}
M.-J. Hwang, R.~Yamamoto, E.~Song, and J.-M. Kim, ``{TTS}-by-{TTS}:
  {TTS}-driven data augmentation for fast and high-quality speech synthesis,''
  \emph{arXiv preprint arXiv:2010.13421}, 2020.

\bibitem{xvector}
D.~Snyder, D.~Garcia-Romero, G.~Sell, D.~Povey, and S.~Khudanpur, ``X-vectors:
  Robust dnn embeddings for speaker recognition,'' in \emph{ICASSP}.\hskip 1em
  plus 0.5em minus 0.4em\relax IEEE, 2018, pp. 5329--5333.

\bibitem{multi-melgan}
G.~Yang, S.~Yang, K.~Liu, P.~Fang, W.~Chen, and L.~Xie, ``{M}ulti-band
  {M}el{GAN}: Faster waveform generation for high-quality text-to-speech,'' in
  \emph{Proc. SLT}, Jan. 2021.

\bibitem{gan}
I.~Goodfellow, J.~Pouget-Abadie, M.~Mirza, B.~Xu, D.~Warde-Farley, S.~Ozair,
  A.~Courville, and Y.~Bengio, ``Generative adversarial nets,'' in \emph{Proc.
  NIPS}, Dec. 2014, pp. 2672--2680.

\bibitem{dcnn}
F.~Yu and K.~Vladlen, ``Multi-scale context aggregation by dilated
  convolutions,'' in \emph{Proc. ICLR}, May 2016.

\bibitem{leakyrelu}
A.~L. Maas, A.~Y. Hannun, and A.~Y. Ng, ``Rectifier nonlinearities improve
  neural network acoustic models,'' in \emph{Proc. ICML}, June 2013, pp. 3--11.

\bibitem{multi-wn-vocoder}
T.~Okamoto, K.~Tachibana, T.~Toda, Y.~Shiga, and H.~Kawai, ``An investigation
  of subband {W}ave{N}et vocoder covering entire audible frequency range with
  limited acoustic features,'' in \emph{ICASSP}.\hskip 1em plus 0.5em minus
  0.4em\relax IEEE, 2018, pp. 5654--5658.

\bibitem{multi-fftnet}
T.~Okamoto, T.~Toda, Y.~Shiga, and H.~Kawai, ``Improving {FFTN}et vocoder with
  noise shaping and subband approaches,'' in \emph{SLT}.\hskip 1em plus 0.5em
  minus 0.4em\relax IEEE, 2018, pp. 304--311.

\bibitem{multi-wavernn}
C.~Yu, H.~Lu, N.~Hu, M.~Yu, C.~Weng, K.~Xu, P.~Liu, D.~Tuo, S.~Kang, G.~Lei
  \emph{et~al.}, ``Dur{IAN}: Duration informed attention network for multimodal
  synthesis,'' \emph{arXiv preprint arXiv:1909.01700}, 2019.

\bibitem{plda_2006}
S.~Ioffe, ``Probabilistic linear discriminant analysis,'' in \emph{ECCV}.\hskip
  1em plus 0.5em minus 0.4em\relax Springer, 2006, pp. 531--542.

\bibitem{plda_2007}
S.~J. Prince and J.~H. Elder, ``Probabilistic linear discriminant analysis for
  inferences about identity,'' in \emph{ICCV}.\hskip 1em plus 0.5em minus
  0.4em\relax IEEE, 2007, pp. 1--8.

\bibitem{m2voc2021}
Q.~Xie, X.~Tian, G.~Liu, K.~Song, L.~Xie, Z.~Wu, H.~Li, S.~Shi, H.~Li, F.~Hong,
  H.~Bu, and X.~Xu, ``The multi-speaker multi-style voice cloning challenge
  2021,'' in \emph{ICASSP}.\hskip 1em plus 0.5em minus 0.4em\relax IEEE, 2021.

\bibitem{adam}
D.~P. Kingma and J.~L. Ba, ``Adam: {A} method for stochastic optimization,'' in
  \emph{Proc. ICLR}, May 2015.

\bibitem{demo}
\BIBentryALTinterwordspacing
Y.-C. Wu, \emph{{RDS} demo}, Accessed: 2021. [Online]. Available:
  \url{https://bigpon.github.io/RelationalDataSelection_demo/}
\BIBentrySTDinterwordspacing

\end{thebibliography}

\end{document}